\documentstyle[12pt]{article}
\newcommand{\mibitem}[1]{\bibitem{#1}}

\newcommand{\nn}{\nonumber}
\newcommand{\be}{\begin{equation}}
\newcommand{\ee}{\end{equation}}
\newcommand{\ba}{\begin{eqnarray}}
\newcommand{\ea}{\end{eqnarray}}
\newcommand{\bastar}{\begin{eqnarray*}}
\newcommand{\eastar}{\end{eqnarray*}}

\newcommand{\M}{{\cal M}}

\newcommand{\cS}{{\cal S}}

\newcommand{\X}{{\cal X}}

\newcommand{\cL}{{\cal L}}

\begin{document}

\begin{titlepage}
\begin{flushright}
CERN-TH.7512/94\\
UU-ITP 18-1994 \\
hep-th/9412023
\end{flushright}

\vskip 0.3truecm

\begin{center}
{ \large \bf
ON THE CHARACTERIZATION OF
\\ \vskip 0.2cm
CLASSICAL DYNAMICAL SYSTEMS USING
\\ \vskip 0.2cm
SUPERSYMMETRIC NONLINEAR $\bf \sigma$-MODELS
\\  }
\end{center}

\vskip 0.5cm

\begin{center}
{\bf Antti J. Niemi $^{*\ddagger}$ } \\
\vskip 0.3cm
{\it Department of Theoretical Physics, Uppsala University
\\ P.O. Box 803, S-75108, Uppsala, Sweden \\}
\vskip 0.3cm
and \\
\vskip 0.3cm
{\bf Kaupo Palo $^{*}$ } \\
\vskip 0.3cm
{\it Theory Division, CERN \\
CH-1211 Geneva 23, Switzerland \\ }
\vskip 0.0cm
and \\
\vskip 0.0cm
{\it Institute of Physics, Estonian Academy of Sciences \\
142 Riia St., 202 400 Tartu, Estonia }

\end{center}

\vskip 0.6cm

\rm
\noindent
We construct a two dimensional nonlinear $\sigma$-model that
describes the Hamiltonian flow in the loop space of a classical
dynamical system. This model is obtained by equivariantizing
the standard N=1 supersymmetric nonlinear $\sigma$-model by the
Hamiltonian flow. We use localization methods to evaluate the
corresponding partition function for a general class of integrable
systems, and  find relations that can be viewed as generalizations
of standard relations in classical Morse theory.

\vfill

\begin{flushleft}
\noindent
\rule{5.1 in}{.007 in}\\
$^{*}${\small E-mail: $~~$ {\small \bf
niemi@tethis.teorfys.uu.se} $~~~~$ {\bf
palo@nxth04.cern.ch} \\
$^{\ddagger}$ Work supported by NFR Grant F-AA/FU 06821-308 and by \\
\hskip 0.3cm G{\"o}ran Gustafsson's Stiftelse}

\end{flushleft}

\end{titlepage}

\vfill\eject

\baselineskip 0.65cm

Originally quantum field theories were introduced to
understand particle physics, but the techniques
have also found extensive applications in statistical
mechanics and condensed matter physics. Very recently,
quantum field theoretical methods have also been applied
to study aspects of {\it classical} Hamiltonian dynamics
where many important problems remain unsolved, in particular
in the case  of chaotic systems.

In applications to classical dynamical systems, quantum
field theories appear to be particularly effective when one
investigates issues such as counting the number of certain
classical trajectories. Such counting problems can be usually
formulated by properly generalizing the Morse theory \cite{milnor}
to infinite dimensions.  These infinite dimensional versions
are usually ill-defined, but difficulties can be at least
partially cured using a field theory formulation.

One of the most actively studied problems in classical Hamiltonian
dynamics is the Arnold conjecture \cite{arnold}, \cite{hofer} that
estimates a lower bound for the number of periodic solutions to
Hamilton's equations of motion. We consider a compact symplectic
manifold $\M$  with local coordinates $\phi^\mu$ and Poisson bracket
\[
\{ \phi^\mu , \phi^\nu \} ~=~ \omega^{\mu\nu}(\phi)
\]
We assume that $H(t,\phi) : \ {\rm R} \times \M \to {\rm R}$
is a smooth, in general explicitly time dependent
({\it i.e.} energy non-conserving) and time periodic Hamiltonian,
\[
H(t,\phi) ~=~ H(t+T,\phi)
\]
and we are interested in $T$-periodic
solutions to Hamilton's equations of motion
\be
\partial_t \phi^\mu ~=~ \{ \phi^\mu , H \} ~=~ \omega^{\mu\nu}
\partial_\nu H ~ \equiv~
\X^\mu_H
\label{hamilton}
\ee
Such periodic solutions are of fundamental importance in many problems:
For example, they appear in the old Bohr quantization
rule. Furthermore, the dynamics of a classically integrable system
is entirely determined by periodic trajectories restricted
on invariant torii. The Gutzwiller trace formula which is the only
effective tool presently available to study quantum chaos, also
approximates the energy spectrum using periodic classical trajectories.

If the solutions to (\ref{hamilton}) are
non-degenerate, the Arnold conjecture
states that their number is bounded from below by the sum of
the Betti numbers $B_k = {\rm dim} H^k (\M), ~ 0 \leq k
\leq {\rm dim} \M$,
\be
\#\{contractible ~ T\!-\!periodic ~ solutions\} ~\geq~ \sum_{k}
{\rm dim}
\
H^k(\M)
\label{arnold}
\ee
Obviously, this conjecture can be viewed as
a natural generalization of the
familiar finite dimensional Morse
inequality that estimates the number of
critical points of a smooth real-valued function on $\M$
\cite{milnor}, \cite{witten1}. Until now it
has only been proven in special
cases \cite{hofer}, most notably for
symplectic manifolds $\M$ such that the
integral of the symplectic two-form $\omega$ over
every sphere $S^2 \subset
\M$ vanishes,
\be
[\omega] \pi_2(\M)  = 0 ~ \Leftrightarrow ~ \int\limits_{S^2} \omega
{}~=~ 0 ~~~~ {\rm for ~ all ~~}
S^2 \subset \M
\label{S2}
\ee
In particular, for the symplectic one-form $\vartheta_\mu$ such that
$\omega = d\vartheta$ this implies that
\be
\oint\limits_{\partial D} \vartheta_\mu d\phi^\mu ~=~ \int\limits_D
\omega
\label{dirac}
\ee
is independent of the disc $D$ which is bounded by the loop $
\partial D: ~\phi^\mu \to \phi^\mu(t)$. $~$ The condition (\ref{S2})
is very restrictive: The volume-form of a compact $2n$-dimensional
symplectic manifold $\M$ is proportional to $\omega^n$, hence the
cohomology class $[\omega]$ must be nontrivial in $H^2(\M,{\rm R})$.
However if $\pi_1(\M) = 0$, the homomorphism $\pi_2(\M) \mapsto
H_2(\M,{\rm Z})$ is an isomorphism. For (\ref{S2}) to be valid,
we should then consider manifolds with $\pi_1(\M) \neq {0}$ such as the
torus $T^2 = S^1 \times S^1$. At the moment only partial results
are known for manifolds with $[\omega] \pi_2(\M) \neq {0}$ \cite{hofer}.

\vskip 0.2cm

In a series of papers Floer \cite{floer} (for a review see
\cite{hofer}) formulates the Arnold
conjecture in terms of infinite dimensional Morse theory. He uses
the classical action
\be
S_{cl}(\phi) ~=~ \int\limits_0^T \{ \vartheta_\mu {\dot \phi}^\mu -
H(t,\phi)\} dt
\label{classical}
\ee
to define a gradient flow in the space  $\cL\M$ of closed loops
$\phi^\mu (0) = \phi^\mu (T)$,
\be
{ \partial \phi^\mu \over \partial s} ~=~ - g^{\mu\nu} {\delta \over
\delta
\phi^\nu} S_{cl}(\phi)
\label{grad1}
\ee
Here $s$ parametrizes the flow of the loops, and $g_{\mu\nu}$ is a
Riemannian metric which is consistent
with an almost complex structure ${I^\mu}_\nu$ on $\M$,
\[
g^{\mu\lambda}\omega_{\lambda\nu} = {I^\mu}_\nu
\]
\[
{I^\mu}_\lambda
{I^\lambda}_\nu ~=~ -{\delta^\mu}_\nu
\]
Explicitly, (\ref{grad1}) is
\be
{ \partial \phi^\mu \over \partial s} + {I^\mu}_\nu { \partial
\phi^\nu
\over \partial t} + g^{\mu\nu} \partial_\nu H(t,\phi) ~=~ 0
\label{grad2}
\ee
and this equation is defined on the cylinder $R \times S^1$ with local
coordinates $s$ and $t$.

Floer uses the fact that in the nondegenerate case, the bounded
orbits
of (\ref{grad2}) tend asymptotically to the $T$-periodic solutions of
Hamilton's equations of motion (\ref{hamilton}). He  applies
Fredholm theory to construct a chain complex generated by these
$T$-periodic solutions, and shows that the homology of this complex
is {\it independent} of the Hamiltonian $H(t,\phi)$. This implies in
particular, that it defines an invariant of the underlying manifold.
By continuing his complex to the classical Morse complex defined by
the gradient flow
\[
\dot\phi^\mu ~=~ - g^{\mu\nu} \partial_\nu H
\]
Floer then  proves  the non-degenerate Arnold
conjecture (\ref{arnold}) on a symplectic manifold $\M$ subject to
the condition (\ref{S2}),  by showing that his homology group is a
model of the singular homology of the manifold $\M$.

Subsequently, Witten \cite{witten2} formulates Floer homology in terms
of a relativistic, $1+1$-dimensional topological nonlinear
$\sigma$-model.
He uses the fact that since the Floer
homology is  independent of the
Hamiltonian $H$, one should be able to describe it  by entirely
ignoring the Hamiltonian. Instead of (\ref{grad2}),
we then have its $H=0$
version
\be
{ \partial \phi^\mu \over \partial s} + {I^\mu}_\nu { \partial
\phi^\nu
\over \partial t}  ~=~ 0
\label{noH}
\ee
Witten constructs his topological $\sigma$-model so that
the corresponding path integral localizes to the solutions of
this equation. He then relates the number
of these solutions to the Floer
homology, by showing that the space of quantum ground states of his
$\sigma$-model coincides with the Floer group.

Obviously, it would be very interesting to develop
quantum field theory techniques of \cite{witten2} further, so that one
could actually analyze  the $H$-dependent properties of
the space of solutions to
(\ref{hamilton}).  Such a quantum field theory
approach to classical dynamical
systems could prove  most useful in
a number of problems. In particular, it
could develop into a new application of field theory
techniques that might be useful in the investigation of
classically chaotic dynamical systems.

An attempt to formulate classical mechanics using
quantum mechanical path
integrals has ben discussed in \cite{gozzi}. However, the
approach used there is
quite different from that developed in \cite{witten2}.

Recently, Sadov \cite{sadov} has proposed
a very interesting  approach to extend
the methods in \cite{witten2} to include
a nontrivial Hamiltonian flow. He
constructs a nonrelativistic $1+1$ dimensional $\sigma$-model based on the
topological model introduced in \cite{witten2}. However, the
model he constructs is quite elaborated, and until now there
has been very little progress in analyzing its consequences to
classical dynamical systems.

The purpose of the present Letter is to investigate, how path integral
techniques applied to $1+1$ dimensional $\sigma$-models could be used
to study the space of classical solutions of a  Hamiltonian
system. Instead of the topological $\sigma$-model discussed
in \cite{witten2},
\cite{sadov}, we shall here consider
a somewhat simpler construction which is
based on the conventional $N=1$ supersymmetric
non-linear $\sigma$-model. 	In
this way we  find results that extend aspects
of standard  Morse theory to the
loop space of classical solutions.

In order to apply the $N=1$ supersymmetric $\sigma$-model to
describe the space of solutions to (\ref{hamilton}), we first need
to present an appropriate construction of its action
\cite{palo}.  For this, we
introduce two commuting fields $\phi^\mu(x^\pm)$,
$F_\mu(x^\pm)$ and two anticommuting
fields $\eta^\mu(x^\pm)$, $\bar\eta_\mu
(x^\pm)$, where $x^\pm = s\pm t$ are coordinates on the cylinder $S^1
\times R^1$. We interpret
$\phi^\mu $ and $\bar\eta_\mu $ as
coordinates on a {\it supertangent bundle}
$\cS^*\M$, and identify the variables $\eta^\mu $ and  $F_\mu$ as the
corresponding  basic one-forms, $\eta^\mu \sim d\phi^\mu$ and  $F_\mu
 \sim d\bar\eta_\mu$. The nilpotent exterior derivative $d$ on the
exterior algebra in the space of maps from the cylinder to $\cS^*\M$ is
\[
d ~=~ \int dx^+ dx^- ( \eta^\mu \frac{\partial}{\partial \phi^\mu} +
F_\mu \frac{\partial}{ \partial
\bar\eta_\mu})
\]
(In the following we will usually not write explicitly integrals over
$x^\pm$.) We also introduce the interior multiplication operator
along the vector field $(\partial_- \phi^\mu , \partial_-
\bar\eta_\mu)$,
\[
i_- ~=~ \int \partial_- \phi^\mu i_\mu + \partial_- \bar\eta_\mu
\pi^\mu
\]
where $i_\mu$ and $\pi^\mu$ is the basis of contractions dual to
$\eta^\mu \sim
d\phi^\mu$ and $F_\mu \sim d\bar\eta_\mu$ respectively {\it i.e.}
in the space of two-dimensional fields
\[
i_\mu \eta^\nu ~=~ {\delta_\mu}^\nu \delta (s-s') \delta (t-t')
\]
\[
\pi^\mu F_\nu ~=~ {\delta^\mu}_\nu \delta (s-s') \delta (t-t')
\]
We define the following equivariant exterior derivative
\be
Q_- ~=~ d + i_- ~=~ \eta^\mu \partial_\mu + F_\mu \frac{\partial}
{\partial
\bar\eta_\mu} + \partial_- \phi^\mu i_\mu + \partial_- \bar\eta_\mu
\pi^\mu
\label{Q1}
\ee
(Notice that integration over the cylinder $x^{\pm}$ is implicit
here.) The corresponding (functional) Lie-derivative is
\be
{\cL}_- ~=~ Q_-^2 ~=~ \partial_- \phi^\mu \partial_\mu + \partial_-
\eta^\mu i_\mu + \partial_- F_\mu \pi^\mu + \partial_- \bar\eta_\mu
\frac{\partial}{\partial\bar\eta_\mu}
{}~\equiv~ \partial_-
\label{Q1square}
\ee

In order to relate this to the $\sigma$-model,
we introduce the functional
\be
\Phi ~=~ \Gamma^\rho_{\mu\nu} \pi^\mu \eta^\nu {\bar\eta}_\rho
\label{conj}
\ee
where $\Gamma^\rho_{\mu\nu}$ are components of a (metric) connection
on
$\M$ {\it i.e.} with $g_{\mu\nu}$ a Riemannian metric on $\M$,
\[
\Gamma^\rho_{\mu\nu} ~=~ \frac{1}{2}g^{\rho\sigma} ( \partial_\mu
g_{\sigma\nu}
+ \partial_\nu g_{\sigma \mu} - \partial_\sigma g_{\mu\nu} )
\]
We then conjugate (\ref{Q1}) as follows,
\[
Q_- ~ \to ~ e^{-\Phi} Q_- e^{\Phi} ~=~ Q_- - \{ Q_- , \Phi \} +
\frac{1}{2!} \{ \{ Q_- , \Phi \} , \Phi \} ~+~ ...
\]
This gives for the conjugated $Q_-$
\[
Q_- ~ = ~ \eta^\mu \partial_\mu + (F_\mu + \Gamma^\rho_{\mu\nu}
\eta^\nu {\bar
\eta}_\rho) \frac{\partial}{\partial \bar\eta_\mu } + \partial_-
\phi^\mu i_\mu
\]
\be
+ \{ \partial_- \bar\eta_\mu + \Gamma^\rho_{\mu\nu} F_\rho \eta^\nu -
\partial_- \phi^\nu \Gamma^\rho_{\mu\nu} \bar\eta_\rho + \frac{1}{2}
{R^\rho}_{\mu\lambda\nu} \eta^\nu \eta^\lambda {\bar \eta}_\rho \}
\pi^\mu
\label{Q2}
\ee
where ${R^\rho}_{\mu\lambda\nu}$ is the Riemann curvature tensor on
$\M$. Notice that this conjugation leaves the Lie-derivative
(\ref{Q1square}) invariant,
\be
\cL_- ~\to~ e^{-\Phi} \cL_- e^{\Phi} ~=~ \cL_- ~=~ \partial_-
\label{Lie1}
\ee

Using the metric $g_{\mu\nu}$ we define
\ba
\xi_1 ~&=&~ g^{\mu\nu} F_{\mu} \bar\eta_\nu
\nn
\\
\xi_2 ~&=&~ g_{\mu\nu}\partial_+ \phi^\mu \eta^\nu
\label{xi1}
\ea
and introduce
\[
S_\sigma ~=~ Q_- (\xi_1 + \xi_2) ~=~ \int \{ g_{\mu\nu} \partial_+
\phi^\mu \partial_-
\phi^\nu + g_{\mu\nu}F^\mu F^\nu
\]
\be
+ \ \frac{1}{2} R_{\mu\nu\rho\sigma}
\eta^\sigma \eta^\rho \bar\eta^\mu \bar\eta^\nu
\ - \  \eta^\mu  D^{(+)}_{\mu\nu}  \eta^\nu \ - \ \bar\eta^\mu
D^{(-)}_{\mu\nu}  \bar\eta^\nu
\}
\label{nlsm}
\ee
where we have defined
\[
D^{(\pm)}_{\mu\nu} ~=~ g_{\mu\nu} \partial_\pm
+ \partial_\pm \phi^\rho g_{\mu\sigma} \Gamma^\sigma_{\rho\nu}
\]
We identify (\ref{nlsm}) as the action of the N=1 supersymmetric
non-linear $\sigma$-model. In particular, the present construction
can be viewed as an infinite dimensional version \cite{oma1} of the
Mathai-Quillen formalism \cite{mathai}, \cite{bgv}.
This formalism has been previously applied in the investigation of
Witten's topological $\sigma$-model \cite{wu}, \cite{bt}, however
in a manner different from the
present approach. Notice, that as a $Q_-$ exact
quantity the action has the standard form of a topological action
\cite{BBRT}.

\vskip 0.2cm

The action (\ref{nlsm}) has an obvious $\pm$-symmetry. Indeed, if in
(\ref{Q2}) we exchange $x^+ \leftrightarrow x^-$, $\eta
\leftrightarrow \bar\eta$, $F \leftrightarrow - F$ so that
\[
Q_- ~\to ~ Q_+ ~=~  \bar\eta^\mu \partial_\mu + ( -F_\mu +
\Gamma^\rho_{\mu\nu} \bar\eta^\nu
\eta_\rho  ) {\partial \over \partial  \eta_\mu } + \partial_+
\phi^\mu j_\mu
\]
\be
+ \{ - \partial_+  \eta_\mu + \Gamma^\rho_{\mu\nu} F_\rho
\bar\eta^\nu -
\partial_+ \phi^\nu \Gamma^\rho_{\mu\nu}  \eta_\rho + \frac{1}{2}
{R^\rho}_{\mu\lambda\nu} \bar\eta^\nu \bar\eta^\lambda {  \eta}_\rho
\} \pi^\mu
\label{Q+}
\ee
where $j_\mu$ denotes contraction dual to the one-form
$\bar\eta^\mu$,
and instead of (\ref{xi1}) we define
\[
\bar\xi_1 + \bar\xi_2 ~=~ - g^{\mu\nu} F_{\mu} \eta_\nu
+ g_{\mu\nu}\partial_- \phi^\mu \bar\eta^\nu
\]
we find that we can also represent (\ref{nlsm}) as
\[
S_{\sigma} ~=~ Q_+ ( \bar\xi_1 + \bar\xi_2)
\]
Instead of (\ref{Lie1}) we now find for the corresponding Lie
derivative
\be
Q_+^2 ~=~ \cL_+ ~=~ \partial_+
\label{Lie2}
\ee
and we can identify  $Q_-$ and $Q_+$ as the left and
right chiral generators in the $(1,1)$ world-sheet supersymmetry
algebra  of the nonlinear $\sigma$-model.
This supersymmetry algebra is defined by the relations (\ref{Lie1})
and (\ref{Lie2}) with the addition of
\[
Q_- Q_+ + Q_+ Q_- ~=~ 0
\]

\vskip 0.2cm
We shall now proceed to generalize
the previous construction so, that it can be
used to investigate the solutions of (\ref{hamilton}). For this we first
observe that the zeroes of the $i_\mu ,
\ j_\mu$ components of the vector fields
in (\ref{Q2}), (\ref{Q+}) are solutions to
\be
\partial_{\pm}\phi^\mu ~=~ 0
\label{+-symmetry}
\ee
Consequently we expect  the action (\ref{nlsm})  to
describe the properties of these field configurations. In particular,
if we compare (\ref{+-symmetry}) with the $H=0$ version (\ref{noH}) of
(\ref{grad2}), we conclude that (\ref{+-symmetry})
can be interpreted as the
$H=0$ version of the following Hamiltonian
flow equation in the loop space,
\be
{ \partial \phi^\mu \over \partial s} ~=~ \pm \omega^{\mu\nu} {\delta
\over \delta
\phi^\nu} S_{cl}(\phi)
\label{symp1}
\ee
This is a natural loop space generalization of
(\ref{hamilton}), with the classical
action (\ref{classical}) as a loop space
Hamiltonian function(al): Just like (\ref{grad1}), the
equation (\ref{symp1}) is also a geometric equation that can be
associated with the classical action (\ref{classical}). While the equation
(\ref{grad1}) describes flow between
critical loops of $S_{cl}$, the equation
(\ref{symp1}) describes flow that circulates
around such (isolated) critical
loops.

In component form, (\ref{symp1}) reads
\be
{\partial \phi^\mu \over \partial s} - {\partial \phi^\mu
\over \partial t} + \omega^{\mu\nu}\partial_\nu H (t,\phi) ~\equiv ~
\partial_- \phi^\mu +  \X^\mu_H ~=~ 0
\label{symp2}
\ee
where we have specified the $+$ sign in the {\it r.h.s.} of (\ref{symp1}).
In order to construct the generalization
of the supersymmetric $\sigma$-model
that describes the properties of the classical
trajectories (\ref{hamilton}),
we use (\ref{symp2}) to generalize (\ref{Q1}) to
\be
Q_- ~\to~ Q_S ~=~ \eta^\mu \partial_\mu + F_\mu
\frac{\partial}{\partial
\bar\eta_\mu} + ( \partial_- \phi^\mu - \X^\mu_H ) i_\mu + (
\partial_-
\bar\eta_\mu + \partial_\mu \X^\nu_H \bar\eta_\nu) \pi^\mu
\label{QS1}
\ee
so that the $i_\mu$-component  vanishes on (\ref{symp2})  while the
$\pi^\mu$-component vanishes if $\bar\eta_\mu$ is a Jacobi field of
(\ref{symp2}).

We again introduce the conjugation by (\ref{conj}) which gives
for (\ref{QS1})
\[
Q_S ~=~  \eta^\mu \partial_\mu + (F_\mu + \Gamma^\rho_{\mu\nu}
\eta^\nu {\bar
\eta}_\rho) \frac{\partial}{\partial
\bar\eta_\mu}  + ( \partial_- \phi^\mu
- \X^\mu_H) i_\mu
\]
\be
+ ~ \{ \partial_- \bar\eta_\mu + \partial_\mu \X^\nu_H \bar\eta_\nu
+ \Gamma^\rho_{\mu\nu} F_\rho \eta^\nu -
\partial_- \phi^\nu \Gamma^\rho_{\mu\nu} \bar\eta_\rho +
\Gamma^\nu_{\mu\rho}
\X^\rho_H \bar\eta_\nu +
\frac{1}{2} {R^\rho}_{\mu\lambda\nu} \eta^\nu \eta^\lambda {\bar
\eta}_\rho \}
\pi^\mu
\label{QS2}
\ee
If we now introduce
\be
S_0 ~=~ \int \vartheta_\mu \partial_- \phi^\mu - H + \frac{1}{2}
\eta^\mu
\omega_{\mu\nu} \eta^\nu
\label{action}
\ee
we find that this is a closed quantity with respect to (\ref{QS2}),
\[
Q_S S_0 ~=~ 0
\]
The desired generalization of (\ref{nlsm}) is then
\[
S ~=~ S_0 + Q_S(\xi_1 + \xi_2) ~=~ \int \{ g_{\mu\nu}
\partial_+\phi^\mu
\partial_-\phi^\nu + g_{\mu\nu} F^\mu F^\nu  + \vartheta_\mu
\partial_-\phi^\mu -
g_{\mu\nu} \X^\nu_H  \partial_+ \phi^\mu
- H
\]
\be
- \eta^\mu D^{(+)}_{\mu\nu} \eta^\nu
- \bar\eta^\mu D^{(-)}_{\mu\nu} \bar\eta^\nu
+ \frac{1}{2} \eta^\mu \omega_{\mu\nu} \eta^\nu - \frac{1}{2}
\bar\eta^\mu \Omega_{\mu\nu} \bar\eta^\nu
+ \frac{1}{2} R_{\mu\nu\rho\sigma}
\eta^\sigma \eta^\rho \bar\eta^\mu \bar\eta^\nu \}
\label{eqnlsm}
\ee
where
\be
\Omega_{\mu\nu} ~=~ \partial_\mu (g_{\nu\rho}\X^\rho_H) -
\partial_\nu
(g_{\mu\rho}\X^\rho_H )
\label{Omega}
\ee

In the rest of the present  Letter we shall discuss the
properties of this action. In particular,
we argue that this action describes
the space of classical trajectories  (\ref{hamilton}) in
the sense of (infinite
dimensional) Morse theory. For this, we consider
a special class of Hamiltonians
$H$ with the property, that at each
value of $t$ the metric tensor $g_{\mu\nu}$
is Lie conserved by the Hamiltonian vector field $\X^\mu_H$,
\be
\cL_H g ~=~ \X^\rho_H \partial_\rho g_{\mu\nu} + g_{\mu\rho}
\partial_\nu \X^\rho_{H} + g_{\nu\rho} \partial_\mu \X^\rho_H ~=~ 0
\label{Liemetric}
\ee
The {\it global} existence of such an invariant metric  means that
$H$ - at each time $t$ - is a generator in the Lie algebra of an
isometry group ${\cal G}$ of $g_{\mu\nu}$  that acts canonically on
$\M$. Without loss of generality, we may view $H$ as a
($t$-dependent) $U(1)$-generator in the Cartan subalgebra of this
isometry group.

For this general class of Hamiltonians, the action (\ref{eqnlsm})
admits a $Q_S$-supersymmetry, as  a direct consequence of the fact
that with (\ref{Liemetric}) the  Lie derivative
\be
\cL_H ~=~ \X^\mu_H \frac{\partial}{\partial \phi^\mu}
+ \eta^\mu \partial_\nu \X^\mu_H i_\mu - \partial_\mu \X^\nu_H
\bar\eta_\nu \frac{\partial}{\partial \bar\eta_\mu} - F_\nu
\partial_\mu \X^\nu_H \pi^\mu
\label{covlie}
\ee
transforms the variables $\phi^\mu$, $\eta^\mu$, $F_\mu$ and
$\bar\eta_\mu$ in a generally covariant manner. Hence
\be
Q_S S ~=~ Q_S S_0 + Q_S^2 (\xi_1 + \xi_2) ~=~  Q_S S_0 + (\partial_-
+
\cL_H) (\xi_1 + \xi_2) ~=~ 0
\label{susyalg}
\ee
We observe that if we attempt to
construct the corresponding generalization
of the other supersymmetry generator (\ref{Q+}) we find that the
action (\ref{eqnlsm}) fails to be invariant under a full $(1,1)$
supersymmetry: The inclusion of a nontrivial Hamiltonian flow subject
to (\ref{Liemetric}) explicitly breaks the
$(1,1)$ supersymmetry of the action
(\ref{nlsm}) down to a chiral $(1,0)$ supersymmetry of
(\ref{eqnlsm}). This breaking of
the  $(1,1)$ supersymmetry  can be understood
as follows. The $(+,\eta)$ chiral sector
of the action (\ref{eqnlsm}) contains
the symplectic one-form $\vartheta_\mu$ and the corresponding symplectic
two-form $\omega_{\mu\nu} = \partial_\mu \vartheta_\nu - \partial_\nu
\vartheta_\mu$. The $(-,\bar\eta)$ sector of the action contains
the one-form $\theta_\mu = g_{\mu\nu} \X^\nu_H$ and the corresponding
two-form $\Omega_{\mu\nu}$ (\ref{Omega}) in a similar fashion.
If we define the Hamiltonian
\[
K ~=~ \frac{1}{2} g_{\mu\nu} \X^\mu_H \X^\nu_H ~\equiv~ \frac{1}{2}
g^{\mu\nu} \theta_\mu \theta_\nu
\]
we find that the pair $(H,\omega)$ and $(K,\Omega)$ determines a
bi-hamiltonian pair in the sense that the corresponding classical
equations (\ref{hamilton}) coincide,
\[
\Omega_{\mu\nu} \dot\phi^\nu ~=~ \partial_\mu K ~=~ \Omega_{\mu\nu}
\omega^{\nu\rho} \partial_\rho H
\]
However, since only the Hamiltonian $H$ of the $(+,\eta)$ chiral
sector
appears in (\ref{eqnlsm}) we also understand why the original $(1,1)$
supersymmetry of (\ref{nlsm}) must be broken down to a chiral $(1,0)$
supersymmetry.

{}From the previous observation, we in particular conclude that the action
(\ref{eqnlsm})  naturally incorporates the bi-hamiltonian
structure which is characteristic to integrable models.

If in addition of (\ref{Liemetric}) we also assume that the
symplectic one-form $\vartheta$ is $H$-invariant \cite{blau}
\be
\cL_H \vartheta ~=~ 0 ~~ \Rightarrow ~~ H ~=~ \X^\mu_H \vartheta_\mu
{}~=~
g^{\mu\nu} \vartheta_\mu \theta_\mu
\label{blau}
\ee
we can write (\ref{eqnlsm}) in a functional form which is entirely
$\pm$-symmetric,
\[
S ~=~  \int \{ g_{\mu\nu} \partial_+\phi^\mu
\partial_-\phi^\nu + g_{\mu\nu} F^\mu F^\nu  + \vartheta_\mu
\partial_-\phi^\mu -
\theta_\mu \partial_+ \phi^\mu - g^{\mu\nu} \vartheta_\mu \theta_\mu
\]
\be
- \eta^\mu D^{(+)}_{\mu\nu} \eta^\nu
- \bar\eta^\mu D^{(-)}_{\mu\nu} \bar\eta^\nu
+ \frac{1}{2} \eta^\mu \omega_{\mu\nu} \eta^\nu - \frac{1}{2}
\bar\eta^\mu \Omega_{\mu\nu} \bar\eta^\nu
+ \frac{1}{2} R_{\mu\nu\rho\sigma}
\eta^\sigma \eta^\rho \bar\eta^\mu \bar\eta^\nu \}
\label{eqnlsms}
\ee
In this form, we can readily generalize it to {\it any}  pair of
symplectic structures $(\vartheta, \omega)$ and $(\theta, \Omega)$.
This could be useful in the investigation of bi-Hamiltonian structures.
In particular, we can also write this action  in the form of  a
topological quantum field theory of the cohomological type
\cite{BBRT},
\[
S ~=~ Q_S ( \vartheta_\mu \eta^\mu + \xi_1 + \xi_2)
\]

\vskip 0.3cm
Consider a world-sheet Lorentz transformation
\[
x^{\pm} \to e^{\pm \theta} x^{\pm}
\]
in (\ref{eqnlsm}).  If
we implement this transformation in the action
(\ref{eqnlsm}), we find that its only effect is
to scale the symplectic two-form
\be
\omega_{\mu\nu} ~ \rightarrow ~ e^{-\theta} \omega_{\mu\nu}
\label{lorentz1}
\ee
In the quantum theory, consistency demands that exponential of ($i$
times) the integral of the symplectic one-form over a closed loop
$\partial D$ in $\M$ must be independent of a disk $D$ which is
bounded by the loop; see (\ref{dirac}). This implies in particular,
that the integral of $\omega$ over any $S^2 \subset \M$ must be an
integer multiplet of $2\pi$. Consequently (\ref{lorentz1}) implies
that
unless the additional condition (\ref{S2}) is satisfied, the
world-sheet Lorentz invariance will be broken. Moreover, even if the
condition (\ref{S2}) is satisfied only expectation values of
operators that are independent of $\omega$ can be invariant under
world-sheet Lorentz transformations.

Indeed, the action (\ref{eqnlsms}) can be represented in a
world-sheet Lorentz-covariant form in a very suggestive manner: For
this
we introduce 1+1 dimensional $\gamma$-matrices in terms of the Pauli
matrices by $\gamma^0 = \sigma^2$, $\gamma^1 = i \sigma^1$ and define
a two component spinor by
\[
\psi^\mu ~=~ \left( \begin{array}{c} \eta^\mu \\ \bar\eta^\mu
\end{array} \right)
\]
and as usual,
\[
\bar\psi^\mu ~=~ \psi^{\mu T} \gamma^0
\]
We also define the left and right U(1) gauge fields
\[
A_{+\mu} ~=~ \vartheta_\mu
\]
\[
A_{-\mu} ~=~ - g_{\mu\nu}
\X^\nu_H
\]
and corresponding field strength tensors
\[
F_{+\mu\nu} ~=~ \partial_\mu A_{+\nu} - \partial_\nu A_{+\mu} ~=~
\omega_{\mu\nu}
\]
\[
F_{-\mu\nu} ~=~ \partial_\mu A_{-\nu} - \partial_\nu A_{-\mu} ~=~
\Omega_{\mu\nu}
\]
With $h^{ij}$ the world-sheet Lorentz metric in a light-cone basis
$(i,j = \pm)$ we can then write the action (\ref{eqnlsms}) in
the Lorentz-covariant form
\[
S ~=~ \int \{ g_{\mu\nu} h^{ij} \partial_i \phi^\mu \partial_j
\phi^\nu + g_{\mu\nu} F^\mu F^\nu + h^{ij} A_{i\mu} \partial_j
\phi^\mu - h^{ij}g^{\mu\nu}A_{i\mu}A_{j\nu}
\]
\[
+ \frac{1}{8} R_{\mu\nu\rho\sigma}\bar\psi^\rho \psi^\mu
\bar\psi^\sigma \psi^\nu - \bar\psi^\mu \gamma^i D_i \psi^\nu
+ \frac{1}{2} \bar\psi^\mu \gamma^i F_{i\mu\nu} \psi^\nu \}
\]
Notice that this is also generally covariant on the target manifold
$\M$.

A similar construction also applies to the original action
(\ref{eqnlsm}) with an arbitrary $H$, except that now we can not
identify $H = h^{ij}g^{\mu\nu}A_{i\mu}A_{j\nu}$.

\vskip 0.2cm
In order to relate our action and the space of
classical solutions to (\ref{hamilton}), we  now explicitly evaluate the
path integral
\be
Z ~=~ \int [d\phi] [dF] [d \eta] [d \bar\eta] \exp \{ i S \}
\label{integral1}
\ee
for the general class of Hamiltonians (\ref{Liemetric}) using localization
techniques \cite{bt} with $S$ the $Q_S$-supersymmetric action
(\ref{eqnlsms}). For this we use the fact
\cite{oma2} that (\ref{integral1}) remains invariant under
\[
S ~\to~ S + Q_S \xi
\]
provided $\cL_S \xi = 0$. Using this invariance, we replace in
(\ref{eqnlsm})
\[
\xi_2 ~\rightarrow~ \lambda \xi_2 ~=~ \lambda g_{\mu \nu} \partial_+
\phi^\mu \eta^\nu
\]
where $\lambda$ is a parameter. For the action (\ref{eqnlsms}) this
gives
\[
S + Q_S(\xi_1 + \lambda \xi_2) ~=~  \int \{ g_{\mu\nu}
\partial_+\phi^\mu
\partial_-\phi^\nu + g_{\mu\nu} F^\mu F^\nu  + \vartheta_\mu
\partial_-\phi^\mu -
\lambda g_{\mu\nu} \X^\nu_H  \partial_+ \phi^\mu - H
\]
\be
- \lambda \eta^\mu D^{(+)}_{\mu\nu} \eta^\nu
-  \bar\eta^\mu D^{(-)}_{\mu\nu} \bar\eta^\nu
+ \frac{1}{2} \eta^\mu \omega_{\mu\nu} \eta^\nu - \frac{1}{2}
\bar\eta^\mu \Omega_{\mu\nu} \bar\eta^\nu + \frac{1}{2}
R_{\mu\nu\rho\sigma}
\eta^\sigma \eta^\rho \bar\eta^\mu \bar\eta^\nu \}
\label{nlsm1}
\ee
In order to evaluate the  path integral we first expand  the
fields $\phi^\mu (x^+,x^-)$ and $\eta^\mu(x^+,x^-)$ as follows,
\[
\phi^\mu (x^+, x^-) ~=~ \phi_-^\mu (x^-) + \delta \phi^\mu (x^+,x^-)
\]
\[
\eta^\mu (x^+, x^-) ~=~ \eta_-^\mu (x^-) + \delta \eta^\mu (x^+,x^-)
\]
That is, we separate the part which is independent of $x^+$. The path
integral measure in (\ref{integral1}) is now defined by
\[
[d\phi][d\eta] ~=~ [d\phi_-] [d\eta_-] [d\delta \phi] [d\delta\eta]
\]
We introduce the following change of variables which has unit
Jacobian in the path integral measure
\[
\delta \phi^\mu (x^+, x^-) ~ \rightarrow \frac{1}{\sqrt{\lambda}}
\delta \phi^\mu(x^+,x^-)
\]
\[
\delta \eta^\mu (x^+, x^-) ~ \rightarrow \frac{1}{\sqrt{\lambda}}
\delta \eta^\mu(x^+,x^-)
\]
and set $\lambda \to \infty$. In this limit the only surviving terms
involving $\delta \phi$ and $\delta \eta$ are
\be
\delta \phi^\mu ( - D^{(-)}_{\mu\nu} - \frac{1}{2}\Omega_{\mu\nu} +
\frac{1}{2} R_{\rho\sigma\mu\nu} \eta^\sigma_- \eta^\rho_-)
\partial_+ \delta\phi^\nu
+ \delta\eta^\mu ( - g_{\mu\nu} \partial_+ ) \delta \eta^\nu
\label{fluct1}
\ee
We evaluate the corresponding functional integral which gives
\be
{\det}^{- \frac{1}{2} } || - { D^{ (-)\mu } }_\nu -
\frac{1}{2} {\Omega^\mu}_\nu + \frac{1}{2} {R^\mu}_{\nu\rho\sigma}
\eta^\sigma_- \eta^\rho_- ||
\label{det1}
\ee
Next, we consider the integral over $\bar\eta^\mu$ and $F^\mu$: We
again introduce
\[
{\bar\eta}^\mu (x^+ x^-) ~=~ \bar\eta^\mu_- (x^-) + \delta
\bar\eta^\mu (x^+ x^-)
\]
\[
F^\mu(x^+ x^-) ~=~ F^\mu_- (x^-) + \delta F^\mu(x^+ x^-)
\]
and define the path integral measures accordingly. We
then find that the integral over $\delta\bar\eta^\mu$ and $\delta
F^\mu$ {\it cancels} (\ref{det1}). As a consequence the
original  $D=2$ path integral (\ref{integral1}) reduces to a {\it one
dimensional} path integral with action
\[
S_{D=1} ~=~ \int d\tau \{ g_{\mu\nu}F^\mu F^\nu + \vartheta_\mu
\partial_\tau \phi^\mu - H(\tau, \phi)
+ \frac{1}{2} \eta^\mu \omega_{\mu\nu} \eta^\nu
\]
\be
+ \bar\eta^\mu ( - D_{\tau\mu\nu} - \frac{1}{2} \Omega_{\mu\nu}
+ \frac{1}{2} R_{\mu\nu\rho\sigma}\eta^\rho \eta^\sigma )
\bar\eta^\nu
\}
\label{qm1}
\ee
where the fields now depend on the variable $\tau \equiv x^-$ only,
and in particular $H(\tau, \phi)$ is the average of $H(t,\phi)$ over
$x^+$.

The action (\ref{qm1}) is equivariantly closed with respect to the
one-dimensional version of (\ref{QS2}),
\[
Q_{D=1} ~=~ \eta^\mu { \partial \over \partial \phi^\mu} + (F_\mu +
\Gamma^\rho_{\mu\nu} \eta^\nu \bar\eta_\rho ) \frac{\partial}{
\partial \bar\eta_\mu } + ( \partial_\tau\phi^\mu - \X^\mu_H ) i_\mu
\]
\[
+ \{ \Gamma^\rho_{\mu\nu}F_\rho \eta^\nu - \frac{1}{2}
{R^\rho}_{\mu\sigma\nu} \eta^\nu \eta^\sigma \bar\eta_\rho -
(\partial_\tau \phi ^\nu - \X^\nu_H) \Gamma^\rho_{\mu\nu}
\bar\eta_\rho
+ ( {\delta^\rho}_\mu \partial_\tau + \partial_\mu \X^\rho_H )
\bar\eta_\rho  \} \pi^\mu
\]

\[
Q_{D=1} S_{D=1} ~=~ 0
\]
Consequently, if we consider the following one-dimensional path
integral
\be
Z_{D=1} ~=~ \int [d\phi][dF] [d\eta] [d\bar\eta] \exp\{ i S_{D=1} +
Q_{D=1}  \xi \}
\label{qmpath}
\ee
we conclude  that it is independent of $\xi$ whenever
\be
Q_{D=1}^2 \xi ~=~ \cL_{D=1} \xi ~=~ 0
\label{uusiLie}
\ee
Hence it coincides with the path integral for the action (\ref{qm1}),
obtained by setting $\xi=0$.

For (\ref{uusiLie}), it is sufficient that $\xi$ is
generally covariant. If we select
\be
\xi ~=~ g^{\mu\nu} F_\mu \bar\eta_\nu + {\lambda \over 2} g_{\mu\nu}
(\partial_\tau  \phi^\mu - \X^\mu_H) \eta^\nu
\label{qmxi1}
\ee
where $\lambda$ is a parameter, we conclude that (\ref{qmpath}) is
independent of $\lambda$ and we can take $\lambda\to\infty$. If we
assume that the $T$-periodic solutions to (\ref{hamilton}) (where
$T$ now denotes a
period in $\tau = x^-$) are non-degenerate.
With $S_{cl}(\phi)$  the corresponding classical action
(\ref{classical}), we find that (\ref{qmpath}) localizes to critical
points of $S_{cl}$,
\be
Z ~=~ Z_{D=1} ~=~ \sum_{\delta S_{cl} = 0} {\rm sign}( \det ||
{\delta^2 S_{cl}  \over
\delta \phi^\mu \delta \phi^\nu } || ) \exp \{ i S_{cl} \}
\label{GBC}
\ee
Obviously, we can view this as an equivariant loop space generalization
of the quantity that appears in the Gauss-Bonnet-Chern
theorem in classical Morse theory \cite{milnor}, \cite{witten1},
\cite{bgv}. Hence we conclude in particular, that our field
theory (\ref{eqnlsm}) describes the space of classical solutions
to (\ref{hamilton}) in the sense
of equivariant loop space Morse theory.

We can relate (\ref{GBC}) to the
topology of $\M$ as follows: If we
select
\[ \xi ~=~ g^{\mu\nu} F_\mu \bar\eta_\nu + {\lambda \over 2}
g_{\mu\nu} \partial_\tau \phi^\mu \eta^\nu
\]
we find in the $\lambda \to \infty$ limit a finite dimensional
integral
over $\M$,
\be
Z ~=~ Z_{D=1} ~=~  \int\limits_{\M} \exp\{ - i T ( H + \frac{1}{2}
\eta^\mu \omega_{\mu\nu} \eta^\nu ) \}
{\rm Pf} [ \ \frac{1}{2} ( {\Omega^\mu}_{\nu} +
{R^\mu}_{\nu\rho\sigma}
\eta^\rho \eta^\sigma ) ]
\label{PH}
\ee
Here we recognize a combination of the equivariant Chern character
and the equivariant Euler class for the Hamiltonian $H$ \cite{bgv}.
Consequently we identify (\ref{PH}) as an equivariant version of the
quantity that appears in the Poincar{\'e}-Hopf theorem in classical
Morse theory \cite{milnor}, \cite{witten1}, \cite{bgv}. Notice in
particular that contrary to (\ref{GBC}), this result can also be used
if the set of critical trajectories of $S_{cl}$ is degenerate.

We observe that if we set
$H= \vartheta =0$ so that our action (\ref{eqnlsm})
reduces to the action (\ref{nlsm}) of the standard N=1 supersymmetric
nonlinear $\sigma$-model, (\ref{PH}) reduces to the Euler
characteristic of the manifold $\M$ consistent with the results in
\cite{witten3}.

\vskip 0.2cm
Combining (\ref{GBC}) and (\ref{PH}) we find
\[
\sum_{\delta S_{cl} = 0} {\rm sign}( \det || {\delta^2 S_{cl}  \over
\delta \phi^\mu \delta \phi^\nu } || ) \exp \{ i S_{cl} \}
\]
\be
= ~ \int\limits_{\M} \exp\{ - i T ( H + \frac{1}{2}
\eta^\mu \omega_{\mu\nu} \eta^\nu ) \}
{\rm Pf} [ \ \frac{1}{2} ( {\Omega^\mu}_{\nu} +
{R^\mu}_{\nu\rho\sigma}
\eta^\rho \eta^\sigma ) ]
\label{morse1}
\ee
and in particular our generalization of
the supersymmetric $\sigma$-model
describes the space of solutions
to (\ref{hamilton}) in a manner which
generalizes the familiar Gauss-Bonnet-Chern and
Poincar{\'e}-Hopf theorems of
classical Morse theory to equivariant loop space
context. Indeed, if we take $T
\to 0$, we obtain the standard result
\[
\sum\limits_{d H = 0 }  {\rm sign}( \det || {\partial^2 H  \over
\partial \phi^\mu \partial \phi^\nu } || )
\]
\be
= \int\limits_{\M} {\rm Pf} [ \ {R^\mu}_{\nu\rho\sigma} \eta^\rho
\eta^\sigma \ ]
{}~\equiv~ \sum\limits_{k} (-)^k {\rm dim} \ H^k(\M)
\label{morse2}
\ee

The Hamiltonians (\ref{Liemetric}) are examples of {\it perfect Morse
functions} \cite{kirwan} which means that their
Morse index {\it i.e.} the number of negative eigenvalues of
$(\partial^2 H)_{\mu\nu}$  is even for every critical point $dH=0$.
Consequently the {\it l.h.s.} of (\ref{morse2})  counts the number of
these critical points. In particular, in
the $T\to 0$ limit (\ref{GBC}) counts the number of solutions to
(\ref{hamilton}).  This is the quantity that appear in the {\it
l.h.s.} of the Arnold conjecture (\ref{arnold}).
Furthermore, since $\M$ admits perfect Morse functions only if all
$H^{2k+1}(\M) = 0$, we conclude from the {\it r.h.s.} of
(\ref{morse2}) that in the $T\to 0$ limit (\ref{PH}) yields the {\it
r.h.s.} in (\ref{arnold}).
Consequently (\ref{morse1}) is consistent with the Arnold conjecture
as $T\to 0$.

This result means in particular, that in some sense the classical
actions with (\ref{Liemetric}) can  be viewed as ``perfect Morse
functionals'' {\it i.e.} as functionals that in a natural fashion
saturate the lower bound of the inequality (\ref{arnold}). However,
since the number of classical solutions to (\ref{hamilton}) depends
on $T$, the definition of a perfect Morse {\it functional} is not
necessarily very useful.

We note that as a consequence of the homogeneity of the
Pfaffian, (\ref{PH}) is manifestly Lorentz-invariant in the sense
that if we introduce the world-sheet Lorentz transformation $x^{\pm}
\to  e^{-\theta} x^{\pm}$ so that  $\omega$ scales according to
(\ref{lorentz1}), the partition function
(\ref{PH}) is independent of $\theta$ since all $\theta$-dependence
from $\omega$ is cancelled by $\theta$-dependence from $\Omega$.
Moreover, if instead of $\M$ we integrate (\ref{PH}) over nontrivial
lower dimensional cycles in $\M$, the $\theta$-dependence appears
only as an overall normalization factor. Hence we may  view such
nontrivial cycles as Lorentz-invariant ``observables'' in our theory,
for manifolds
that satisfy the condition (\ref{S2}). However, we also note that if
all
odd Betti numbers $B_{2k+1}$ vanish and in addition $\omega$
satisfies   (\ref{S2}), we obtain further conditions from the ensuing
requirement that $\pi_1(\M) \neq 0$: Since $H_1(\M)$ is the
Abelianization of $\pi_1(\M)$ \cite{green}, we conclude that the
Abelian part of $\pi_1(\M)$ must vanish. Recently \cite{gompf},
examples of four dimensional symplectic manifolds with $\pi_1 \sim G$
where $G$ is any finitely presented group, have been constructed. For
$G$ that does not admit an Abelian part, we then have $B_1 = B_3 = 0$
and $\pi_2(\M) \neq H_2(\M,Z)$. If in addition $\omega$ satisfies
(\ref{S2}), it might become possible to construct examples of
supersymmetric Lorentz invariant theories based on an equivariant
generalization of the
standard supersymmetry algebra.  However, we note that from the point
of view of
classical dynamical systems, world sheet Lorentz invariance is  not an
important property.

\vskip 0.2cm
Unfortunately, at the moment there is not very much we can say about
(\ref{eqnlsm})  for  a general $t$-dependent Hamiltonian $H(t,\phi)$:
The path integral (\ref{integral1}) is too complicated to be explicitly
evaluated in the general case. However, it is natural to expect that
also in the  general case our theory describes the space of classical
trajectories in the sense of
an infinite dimensional Morse theory. Indeed,
we can still identify a {\it local} supersymmetric structure:

If $H$ does not satisfy the condition
(\ref{Liemetric}),  the action (\ref{eqnlsm}) fails to be
(globally) supersymmetric. However, for {\it any} $H$ we can still find a
local supersymmetry in any neighborhood in $\M$ that does not
include critical points of $H$. Using canonical transformations, we
can always identify $H$  {\it e.g.} with some component
$p_i$ of the canonical momentum in such (Darboux) neighborhoods. If
in these neighborhoods we set $g_{\mu\nu} = \delta_{\mu\nu}$, we find
that the condition  (\ref{Liemetric}) is satisfied. Consequently,
locally the construction of (\ref{eqnlsm})  reflects  supersymmetric
structure and the supersymmetry is broken only due to the fact that
$g_{\mu\nu}$ can not be globally defined.

Indeed, even in the general case the action (\ref{eqnlsm}) is
constructed using
a $(1,0)$ supersymmetry algebra: Instead of the supersymmetry algebra
of (\ref{eqnlsms}),
\[
Q_S^2 ~=~ \partial_- +
\cL_H
\]
\[
Q_S (\partial_- + \cL_H) - (\partial_- + \cL_H) Q_S ~=~ 0
\]
in the general case the construction of the action (\ref{eqnlsm})
uses the isomorphic algebra
\[
Q_S^2 ~=~ \partial_- + \L_H
\]
\[
Q_S (\partial_- + \L_H) -  (\partial_- + \L_H )Q_S ~=~ 0
\]
where
\[
\L_H ~=~ \cL_H + \cL \Gamma^\rho_{\mu\nu}  \bar\eta_\rho \eta^\nu
\pi^\mu
\]
and $\cL \Gamma^\rho_{\mu\nu}$ denotes Lie derivative of the
connection $\Gamma^\rho_{\mu\nu}$ in the {\it original} manifold
$\M$,
\[
\cL \Gamma^\rho_{\mu\nu}   ~=~
\X^\sigma_H \partial_\sigma\Gamma^\rho_{\mu\nu} +
\partial_\nu\X^\sigma_H
\Gamma^\rho_{\mu\sigma}
+ \partial_\mu \X^\sigma_H \Gamma^\rho_{\sigma\nu} -
\Gamma^\sigma_{\mu\nu}
\partial_\sigma \X^\rho_H +
\partial_{\mu\nu} \X^\rho_H
\]
which vanishes if  (\ref{Liemetric}) is satisfied.

\vskip 0.4cm

In conclusion, we have investigated a generalization of the
standard $N=1$ supersymmetric nonlinear $\sigma$-model, obtained
by equivariantizing the $N=1$ model by a nontrivial loop space
Hamiltonian flow. By explicitly evaluating the corresponding
path integral for a class of Hamiltonians we have found,
that our generalization describes properties of a classical
dynamical system in the sense of classical Morse theory. Our approach
suggest that quantum field theoretical methods could be very effective in
analyzing the structure of  classical dynamical systems, and obviously a
generalization of our results to models that do not obey the condition
(\ref{Liemetric}) would be very interesting.

\vskip 1.0cm
A.N. thanks M. Blau, A. Gerasimov,  A. Losev,  O. Viro and S. Wu for
discussions and comments on the manuscript.

\end{document}